\documentclass[preprint,prd,superscriptaddress,amsmath,nofootinbib]{revtex4}

\usepackage{cancel}
\usepackage{graphicx}
\usepackage{color}
\usepackage{amssymb}

\newcommand{\bea}{\begin{eqnarray}}
\newcommand{\eea}{\end{eqnarray}}

\newcommand{\eq}[1]{Eq.~(\ref{#1})}

\begin{document}
\preprint{\vbox{\hbox{CERN-PH-TH-2014-204}}}

\title{1-loop SQCD corrections to the decay of top-squarks to charm and neutralino in the generic MSSM  \bigskip}

\author{Jason Aebischer} \email{aebischer@itp.unibe.ch}
\affiliation{Albert Einstein Center for Fundamental Physics, Institute
  for Theoretical Physics,\\ University of Bern, CH-3012 Bern,
  Switzerland.}
\author{Andreas Crivellin} \email{andreas.crivellin@cern.ch}
\affiliation{CERN Theory Division, CH-1211 Geneva 23, Switzerland}
\author{Christoph Greub}\email{greub@itp.unibe.ch}
\affiliation{Albert Einstein Center for Fundamental Physics, Institute
  for Theoretical Physics,\\ University of Bern, CH-3012 Bern,
  Switzerland.}
\bigskip
\date{\today\bigskip\bigskip}

\begin{abstract}
In this article we calculate the 1-loop supersymmetric QCD (SQCD) corrections
to the decay $\tilde u_1\to c \tilde\chi^0_1$ in the MSSM with generic flavour
structure. This decay mode is phenomenologically important if the mass difference
between the lightest squark $\tilde u_1$ (which is assumed to be mainly stop-like) and the neutralino LSP $\tilde \chi^0_1$ is smaller than the top mass. In such a scenario $\tilde u_1\to t \tilde\chi^0_1$ is kinematically not allowed and searches for $\tilde u_1\to W b \tilde\chi^0_1$ and $\tilde u_1\to c
\tilde\chi^0_1$ are performed. A large decay rate for $\tilde u_1\to c
\tilde\chi^0_1$ can weaken the LHC bounds from $\tilde u_1\to W b \chi^0_1$
which are usually obtained under the assumption ${\rm Br}[\tilde u_1\to W b
\chi^0_1]=100\%$. 

We find the SQCD corrections enhance $\Gamma[\tilde u_1\to c
\tilde\chi^0_1]$ by approximately 10\% if the flavour-violation originates
from bilinear terms. If flavour-violation originates from trilinear terms, the
effect can be $\pm 50\%$ or more, depending on the sign of $A^t$. We note that connecting a theory of SUSY breaking to LHC observables, the shift from the $\overline{\rm DR}$ to the on-shell mass is numerically very important for light stop decays.
\end{abstract}


\maketitle

\section{Introduction}
\label{sec:intro}

Natural Supersymmetry requires light stops in order to cancel the quadratic divergences of the Higgs self-energies involving a top quark while the other supersymmetric partner can be much heavier \cite{Dimopoulos:1995mi,Giudice:2004tc}. Theoretical motivation for light stops also comes from the fact that when starting at a high scale with universal squark masses, the renormalization group evolution (RGE) (known at the two-loop level \cite{Machacek:1984zw,Yamada:1993ga,Martin:1993zk}) generically drive the masses of the third generation squarks to lower values as for example in gravity mediated SUSY breaking scenarios (see for example \cite{Chamseddine:1982jx}). In addition, light stops are also welcome in order to accommodate for the observed relic density within the MSSM \cite{Boehm:1999bj,Ellis:2001nx,Balazs:2004bu,Balazs:2004ae,Ellis:2014ipa,deSimone:2014pda} and to realize baryogensis \cite{Carena:1996wj,Carena:1997ki,Huet:1995sh,Delepine:1996vn,Losada:1998at,Losada:1999tf,Cirigliano:2006dg,Carena:2008vj,Laine:2012jy}. 
\medskip

On the experimental side, the bounds on the stop mass are much weaker than the ones on the other strongly interacting SUSY particles, i.e. squarks of the first two generations \cite{Aad:2014wea,CMS:2014ksa} and the gluino (see for example \cite{Sekmen:2014vsa} for a recent overview of ATLAS and CMS results). Light stops might even be welcome in the light of recent LHC data for $W$-pair production where the observed cross section \cite{CMS:2012daa,ATLAS:2012mec} is slightly above the SM predictions \cite{Campbell:2007ev}. This can be interpreted as a hint for light sleptons, light chargions and/or light stops \cite{Rolbiecki:2013fia,Curtin:2014zua}. However, in order to accommodate the measured Higgs mass of around 125 GeV~\cite{Aad:2012tfa,Chatrchyan:2012ufa} rather heavy stops are required. This tension can be solved if the stop mixing angle is large (or even maximal \cite{Wymant:2012zp}), by promoting the MSSM to the NMSSM/$\lambda$SUSY \cite{Ellwanger:2009dp,Hall:2011aa} or by adding D-term contributions \cite{Batra:2003nj}.
\medskip

Concerning the exclusion limits on stop masses from the LHC, there are still regions in parameter space in which light stops are allowed. If the mass splitting between the stop and the neutralino is bigger than the top mass, the main search channel is $\tilde u_1\to t\tilde\chi^0_1$ and the constraints are stringent \cite{Chatrchyan:2013xna,ATLAS:2013pla}. However, if the mass difference is smaller than $m_{t}$ the limits on the stop mass come from searches for $\tilde u_1\to Wb\tilde\chi^0_1$ and the limits are much weaker \cite{CMS:2013ida,TheATLAScollaboration:2013gha,Aad:2014kra,Aad:2014nra}. If the mass difference between the stop and the neutralino is even smaller than $m_W+m_b$ the limits are obtained from searches for the flavour-changing decay $\tilde u_1\to c\tilde\chi^0_1$ \cite{TheATLAScollaboration:2013aia,CMS:2014yma}.
\medskip

The decay $\tilde u_1\to c\tilde\chi^0_1$ has important experimental implications, both for scenarios with minimal and non-minimal flavour-violation:
\medskip

In the case of minimal flavor violation (MFV) \cite{Chivukula:1987fw,Hall:1990ac,Buras:2000dm,D'Ambrosio:2002ex,Bobeth:2005ck} the decay rate is suppressed leading to a sizable stop decay length, which can be used to determine the flavour structure \cite{Hiller:2008wp,Hiller:2009ii} and is in principle measurable at the LHC \cite{Han:2003qe,Kraml:2005kb,Bornhauser:2010mw}\footnote{If the decay rate for $\tilde u_1\to c\tilde\chi^0_1$ is small, the four body decay $\tilde u_1\to b\tilde\chi^0_1 ff^\prime$ \cite{Boehm:1999tr} (also searched for at the LHC \cite{Aad:2014kra}) can have a significant impact on branching ratio for $\tilde u_1\to c\tilde\chi^0_1$ \cite{Muhlleitner:2011ww}.}. The most plausible scenario with a suppressed stop decay rate $\tilde u_1\to c\tilde\chi^0_1$ is to assume a flavour-blind SUSY breaking mechanism at some high scale $\Lambda$, for example the GUT scale. In this case, flavour off-diagonal elements in the squark mass matrices are induced by the RG for which the decay width has been calculated in Ref.~\cite{Hikasa:1987db} and the finite part of the 1-loop electroweak corrections has been computed in Ref.~\cite{Muhlleitner:2011ww}\footnote{The corresponding corrections for the flavour conserving case were calculated in Ref.~\cite{Kraml:1996kz,Djouadi:1996wt}.}.

\medskip
In the case of non-minimal flavour violation the decay width for $\tilde u_1\to c\tilde\chi^0_1$ can be significantly enhanced since the flavour-changing elements in the up-sector are rather poorly constrained from FCNC processes. It has been noticed in Ref.~\cite{Blanke:2013uia} (see also \cite{Bartl:2012tx,Agrawal:2013kha} for later analysis) that an enhanced branching ratio for $\tilde u_1\to c\tilde\chi^0_1$ can weaken the bounds from $\tilde u_1\to t\tilde\chi^0_1$, for which a branching ration of $100\%$ is commonly assumed in the experimental analysis, allowing for lighter stop masses. We point out that a similar effect occurs concerning the limits extracted from $\tilde u_1\to W b \chi^0_1$ searches. Since $\tilde u_1\to W b \chi^0_1$ is a three body decay, it is kinematically suppressed compared to the two body decay $\tilde u_1\to t\tilde\chi^0_1$. Therefore, already a much smaller amount of flavour violation, as the one necessary to affect the limits from $\tilde u_1\to t\tilde\chi^0_1$, would be sufficient to significantly weaken the limits extracted from $\tilde u_1\to W b \chi^0_1$. This observation is especially interesting taking into account that the bounds on the stop mass from $\tilde u_1\to W b \chi^0_1$ are currently anyway the weakest ones. Therefore, very light stop masses for $m_W<m_{\tilde u_1}-m_{\tilde\chi^0}<m_{t}$ are allowed, especially in the case of non-minimal flavour-violation.
\medskip

In this article we investigate the 1-loop SQCD corrections to $\tilde u_1\to c\tilde\chi^0_1$ in the MSSM\footnote{Even though we refer to the MSSM here, our analysis does not depend on the Higgs sector of the MSSM and thus also applies to non-minimal extensions like the NMSSM and $\lambda$SUSY \cite{Ellwanger:2009dp,Hall:2011aa}.} with generic flavour structure. These $\alpha_s$ corrections are the leading ones in case of non-minimal flavor violation. Furthermore, assuming a flavour-blind SUSY-breaking mechanism at a high scale $\Lambda$ the counting of the loop-effects is as follows: The leading order effect is the one-loop electroweak running from $\Lambda$ to $m_{\rm SUSY}$. To this leading effect the next-to-leading order (NLO) corrections are the two-loop RGE effects~\cite{Machacek:1984zw,Yamada:1993ga,Martin:1993zk} originating from $\alpha_s$ and the one-loop QCD corrections to the decay width at the SUSY scale which we calculate here\footnote{This work was presented at the SUSY conference 2014 \cite{Jason}. During completion of our work the SUSY-QCD corrections to the decay $\tilde t\to c \tilde\chi^0_1$ have been presented for the first time~\cite{Grober:2014aha}. In that paper furthermore a phenomenological analysis including the flavour-changing two-body decay of the lightest stop into a charm quark and the lightest neutralino and its four-body decay into the lightest neutralino, a down-type quark and a fermion pair, has been performed. However, Ref.~\cite{Grober:2014aha} uses a different renormalization scheme than we do.}. 
\medskip

The article is structured as follows: In the next section we establish our conventions and recall the tree-level expression for the decay rate for $\tilde u_1\to c\tilde\chi^0_1$. Sec.~\ref{calculation} describes the calculation as well as the renormalization followed by a numerical analysis \ref{numerics}. Finally we conclude in Sec.~\ref{conclusions}.
\medskip

\section{Conventions and Tree-level decay}

In this section we define our conventions and discuss the tree-level decay width. First, we denote the term in the Lagrangian for the coupling of an up-quark $u_i$ to a up-squark $\tilde u_s$ and a neutralino $\tilde \chi^0_p$ as
\begin{equation}
\tilde u_s^* \bar{\tilde\chi}_p^0 \left[ {\Gamma _{{{\tilde u}_s}{u_i}}^{\tilde \chi _p^0L}{P_L} + \Gamma _{{{\tilde u}_s}{u_i}}^{\tilde \chi _p^0R}{P_R}} \right]u_i\;+\;{\rm h.c.}\,,
\label{squark-quark-neutralino-vertex}
\end{equation}
where $P_L$ and $P_R$ are chiral projectors. For the coupling of quarks to squarks and gluinos we introduce a similar notation:
\begin{equation}
\tilde u_s^* \bar{\tilde g}\left[ {\Gamma _{\tilde u_s u_i }^{\tilde gL}{P_L} + \Gamma _{\tilde u_s u_i }^{\tilde gR}{P_R}} \right]u_i\;+\;{\rm h.c.}\,.
\end{equation}
In the following, we will order the mass eigenstates for the neutralino $p=1-4$ and of the up-squarks $s=1-6$ in increasing order and $u_3$, $u_2$ and $u_1$ correspond to the $t$, $c$ and $u$ quark, respectively. For the neutralino mass matrix we use the convention
\begin{equation}
{\cal M}_{\tilde \chi _{}^0}^{} = \left( {\begin{array}{*{20}{c}}
{{M_1}}&0&{\dfrac{{ - {v_d}{g_1}}}{{\sqrt 2 }}}&{\dfrac{{{v_u}{g_1}}}{{\sqrt 2 }}}\\
0&{{M_2}}&{\dfrac{{{v_d}{g_2}}}{{\sqrt 2 }}}&{\dfrac{{ - {v_u}{g_2}}}{{\sqrt 2 }}}\\
{\dfrac{{ - {v_d}{g_1}}}{{\sqrt 2 }}}&{\dfrac{{{v_d}{g_2}}}{{\sqrt 2 }}}&0&{ - \mu }\\
{\dfrac{{{v_u}{g_1}}}{{\sqrt 2 }}}&{\dfrac{{ - {v_u}{g_2}}}{{\sqrt 2 }}}&{ - \mu }&0
\end{array}} \right)\,,
\end{equation}
with $v=\sqrt{2}m_W/g_2\approx 174 \,$ GeV and $v_u/v_d=\tan\beta$. The up-squark mass term in the Lagrangian is given by 
\begin{equation}
- \left( {\begin{array}{*{20}{c}}
{\tilde u_L^*}&{\tilde u_R^*}
\end{array}} \right){\cal M}_u^2\left( {\begin{array}{*{20}{c}}
{{{\tilde u}_L}}\\
{{{\tilde u}_R}}
\end{array}} \right)\,.
\end{equation}
where both $\tilde u_L$ and $\tilde u_R$ are 3-vectors in flavour space. The squark mass(-squared) matrix is given by
\begin{equation}
\renewcommand{\arraystretch}{2.0}
	{\cal M}_{\tilde{u}}^2 = 
	\left(\begin{array}{cc}		
		{\bf m}_U^{LL2} + v_u^2 {\bf Y}_u{\bf Y}_u^\dagger &  {\bf \Delta}^{uLR}\\
{\bf \Delta}^{uLR\dagger} & {\bf m}_U^{RR2} + v^2{\bf Y}_u^\dagger{\bf Y}_u
	\end{array}	\right) \,,
	\label{up-squark-mass-matrix}
\end{equation}
with 
\begin{equation}
{\bf \Delta}^{uLR}={\bf \Delta}^{uRL\dagger}=- v_u({\bf A}_u + {\bf Y}_u\mu\cot\beta)\,,\qquad {\bf m}_U^{LL2}=V^\dagger{\bf m}_Q^2 V 	\,.
\end{equation}
Here ${\bf A}_u$, ${\bf m}_U^{LL2}$ and  ${\bf m}_U^{RR2}$ are $3\times 3$ matrices in flavour space and we neglected small terms involving electroweak gauge couplings. Here we allowed for complex Yukawa couplings and used $LR$ conventions for them and the $A$-terms \cite{Crivellin:2011jt}. Note that in \eq{up-squark-mass-matrix} the Yukawa couplings and not the quark masses enter which is a relevant difference since we are computing 1-loop SQCD corrections in this article\footnote{The threshold corrections connecting the Yukawa couplings and the quark masses are known to be very large in the down sector \cite{Banks:1987iu,Hall:1993gn,Hempfling:1993kv,Carena:1994bv,Blazek:1995nv,Hamzaoui:1998nu,Carena:1999py,Buras:2002vd,Hofer:2009xb,Crivellin:2010er} and have been computed at the two-loop level \cite{Bednyakov:2002sf,Bauer:2008bj,Noth:2010jy,Crivellin:2012zz}.}. 
\medskip

\eq{up-squark-mass-matrix} is given in the super-CKM basis which we define to be the basis in which the Yukawa couplings of the MSSM superpotential are diagonal, both for quarks and squarks, so that supersymmertry is manifest:
\begin{equation}
	{\bf Y}_u=\left( {\begin{array}{*{20}{c}}
{{Y^{u_1}}}&0&0\\
0&{{Y^{u_2}}}&0\\
0&0&{{Y^{u_3}}}
\end{array}} \right)\,.
\end{equation}
Note that in the literature the super-CKM basis is often defined to be the basis with diagonal quark mass matrices. However, this definition has the disadvantage that the basis changes with every loop order.
\medskip

We diagonalize the full hermetian $6\times 6$ squark mass-squared matrix ${\cal M}_{\tilde{u}}^2$ and the symmetric $4\times 4$ neutralino mass matrix ${\cal M}_{\tilde \chi^0}$ as 
\begin{align}
W^{\tilde u*}_{s^\prime s}({\cal M}_{\tilde u}^2)_{s^\prime t^\prime}W^{\tilde u}_{t^\prime t}=m^2_{\tilde u_s}\delta_{st}\,,\\
Z_N^{p^\prime p} {\cal M}^{\tilde \chi _{}^0}_{p^\prime q^\prime} Z_N^{q^\prime q}=m_{\tilde \chi^0_{p}}\delta_{pq}\,,
\end{align}
where $Z_N$ and $W^{\tilde u}$ are unitary matrices. With these conventions we get for the squark-quark-neutralino couplings in \eq{squark-quark-neutralino-vertex}:
\begin{equation}
\renewcommand{\arraystretch}{2.0}
\begin{array}{l}
\Gamma _{{{\tilde u}_s}{u_i}}^{\tilde \chi _p^0L} = \dfrac{{ - e}}{{\sqrt 2 {s_W}{c_W}}}W_{is}^{\tilde u*}\left( {\dfrac{1}{3}Z_N^{1p}{s_W} + Z_N^{2p}{c_W}} \right) - {Y^{{u_i}*}}W_{i + 3,s}^{\tilde u*}Z_N^{4p}\,,\\
\Gamma _{{{\tilde u}_s}{u_i}}^{\tilde \chi _p^0R} = \dfrac{{2\sqrt 2 e}}{{3{c_W}}}W_{i + 3,s}^{\tilde u*}Z_N^{1p*} - {Y^{{u_i}}}W_{is}^{\tilde u*}Z_N^{4p*}\,.
\end{array}\,,
\end{equation}
and for the squark-quark gluino vertex
\begin{align}
\Gamma _{\tilde u_s^{}u_i^{}}^{\tilde gL} =  - \sqrt 2 {g_s}T_{}^aW_{is}^{\tilde u*}\,,\\
\Gamma _{\tilde u_s^{}u_i^{}}^{\tilde gR} = \sqrt 2 {g_s}T_{}^aW_{i + 3,s}^{\tilde u*}\,.
\end{align}
Here $e$ denotes the electric charge and $s_W\equiv \sin{\theta_W},c_W\equiv\cos{\theta_W}$, where $\theta_W$ is the Weinberg angle. The tree-level decay width of the lightest squark into the LSP and a (massless) charm quark is given by: 
\begin{equation}
\Gamma_0 \left[ {\tilde u_1 \to u_2\tilde \chi _1^0} \right] = \dfrac{{{m_{{{\tilde u}_1}}}}}{{16\pi }}{\left( {1 - \dfrac{{m_{\tilde\chi _1^0}^2}}{{m_{{{\tilde u}_1}}^2}}} \right)^2}\left( {{{\left| {\Gamma _{{{\tilde u}_1}u_2}^{\tilde \chi _1^0L}} \right|}^2} + {{\left| {\Gamma _{{{\tilde u}_1}u_2}^{\tilde \chi _1^0R}} \right|}^2}} \right)\,. \label{tree-level-decay-width}
\end{equation}
If the LSP is mostly bino like, we can further simplify the expression neglecting very small neutralino mixing and small charm Yukawa couplings:
\begin{equation}
\Gamma_0 \left[ {\tilde u_1 \to u_2\tilde \chi _1^0} \right] = \dfrac{{{m_{{{\tilde u}_1}}}}}{{16\pi }}\dfrac{{g_1^2}}{{18}}\left( {1 - \dfrac{{m_{\chi _1^0}^2}}{{m_{{{\tilde u}_1}}^2}}} \right)^2\left( {{{\left| {W_{21}^{\tilde u}} \right|}^2} + 16{{\left| {W_{51}^{\tilde u}} \right|}^2}} \right)\,.
\end{equation}
Note that the decay to a right-handed charm quark is enhanced by a factor 16 which can be traced back to hyper-charges.

\section{Calculation of the SQCD corrections}\label{calculation}

In this section we discuss in detail the calculation of the 1-loop SQCD corrections including our renormalization scheme. Our calculation involves the following steps: 
\begin{enumerate}
  \item Renormalization of the quark sector.
	\item Renormalization of the squark sector.
	\item Calculation of the gluon contributions to the decay width including real emission corrections, i.e. the decay $\tilde u_1\to c \tilde\chi^0_1 g$.
	\item Calculation of the gluino contributions (including the cancellation of ultraviolet (UV) divergences).
\end{enumerate}

We renormalize the fundamental parameters entering the decay width of the stop decay at tree-level, which receive SQCD corrections at the one loop level, in the $\overline{\rm DR}$ scheme. These quantities are
\begin{itemize}
	\item The Yukawa couplings $Y^{u_i}$ of the MSSM superpotential.
	\item The trilinear $A^u_{ij}$ terms.
	\item The bilinear squark mass terms ${\bf m}_U^{LL2}$ and ${\bf m}_U^{RR2}$.
\end{itemize}
We write the bare quantities of the Lagrangian (labeled with a superscript $(0)$) as
\begin{equation}
Y^{u_i(0)}=Y^{u_i}+\delta Y^{u_i}\,,\qquad A^{u(0)}_{ij}=A^{u}_{ij}+\delta A^{u}_{ij}\,,\qquad{\bf m}_{Q,U}^{2(0)}={\bf m}_{Q,U}^2+{\bf \delta m}_{Q,U}^2\,.\label{barvsrenormalized}
\end{equation}
Since we renormalize all quantities in a minimal renormalization scheme,
i.e. the $\overline{\rm DR}$ scheme, ${\bf A}^{u}$, ${\bf m}_{Q,U}^{2}$ and
${\bf Y}^{u}$ are understood to be the renormalized ones in the
$\overline{DR}$-scheme. However, in the decay width of the stop, the on-shell
squark mass also enters. Therefore, a conversion from the on-shell squark mass
to the $\overline{\rm DR}$ is necessary. In addition, the Yukawa couplings have to be related to the measured quark masses of the SM by running and threshold corrections.  

\subsection{Renormalization of the quark sector}

\begin{figure*}[t]
\centering
\includegraphics[width=0.9\textwidth]{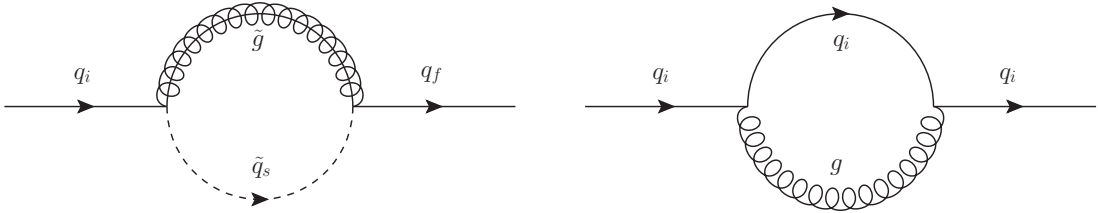}
\caption{Quark self-energy diagrams}
\label{QuarkSE}
\end{figure*}

SQCD corrections to quark masses and Yukawa couplings can be calculated from the quark self-energies (see Fig.~\ref{QuarkSE}). The UV renormalization of the Yukawa couplings (in the $\overline{\rm DR}$-bar scheme) is given by
\begin{equation}
\delta {Y^{u_i}} =  - \dfrac{{{\alpha _s}}}{{2\pi }}\dfrac{1}{\varepsilon }{C_F} {Y^{u_i}} {\mkern 1mu}\,.\label{deltaYquark}
\end{equation}
In our approach we compute only LSZ factors corresponding to flavour-diagonal self-energies, which are also the only UV divergent ones. All other contributions from self-energies can be calculated as one-particle irreducible diagrams \cite{Logan:2000iv,Crivellin:2009ar,Crivellin:2010gw}. Therefore, the LSZ factor for left and right-handed quarks is
\begin{align}
\delta Z_{{u}}^L &= \delta Z_u ^a + \delta Z_{{u_i}}^{Lb}\,,\\
\delta Z_{{u}}^R &= \delta Z_u ^a + \delta Z_{{u_i}}^{Rb}\,.
\end{align}
Here, the superscript $a$ denotes the flavour-independent gluon piece, while the index $b$ refers to the gluino piece whose finite part is in general flavour dependent:
\begin{align}
\delta Z_u ^a &= \dfrac{{{\alpha _s}}}{{4\pi }}{C_F}{\mkern 1mu} (1 - (1 - \xi )){\mkern 1mu} \left[ {\dfrac{1}{{{\varepsilon_{\rm IR}}}} - \dfrac{1}{\varepsilon}} \right]\,,\label{LSZquarkgluon}\\
\delta Z_{{u_i}}^{Lb} &= \Sigma _{ii}^{\tilde g LL}\,,\\
\delta Z_{{u_i}}^{Rb} &= \Sigma _{ii}^{\tilde g RR}\,.
\end{align}
Here $\varepsilon_{\rm IR}$ denotes the dimensionally regularized infrared
(IR) divergence and $\varepsilon$ the UV one while $\Sigma _{ii}^{\tilde g LL}$
and $\Sigma _{ii}^{\tilde g RR}$ are defined in eqs. (\ref{SEdecomposition})
and (\ref{Sigmaabbrev}).

\subsubsection{Threshold corrections}

In order to determine the actual values of the Yukawa couplings we have to
make the connection to the quark masses determined within the SM\footnote{For
  a complete discussion of all one-loop corrections within the MSSM including
  resummation see Ref.~\cite{Crivellin:2011jt}.}. The self-energies with heavy
virtual particles, in our case the one with squarks and gluinos, lead to
threshold corrections modifying the tree-level relation
$v_uY^{u_i}=m_{u_i}$. In order to write down these corrections we decompose the quark self-energies originating from squark-gluino loops as
\begin{equation}
\Sigma _{{u_f}{u_i}}^{\tilde g}\left( {{p^2}} \right) = \Sigma _{fi}^{\tilde g LR}\left( {{p^2}} \right){P_R} + \Sigma _{fi}^{\tilde g RL}\left( {{p^2}} \right){P_L} + \cancel{p}\left( {\Sigma _{fi}^{\tilde g LL}\left( {{p^2}} \right){P_L} + \Sigma _{fi}^{\tilde g RR}\left( {{p^2}} \right){P_L}} \right)\,.\label{SEdecomposition}
\end{equation}
Since in the decay $\tilde u_1\to c \tilde\chi_1^0$ we are dealing with external charm-quarks on the mass-shell it is sufficient to evaluate \eq{SEdecomposition} at vanishing external momenta, i.e. neglecting finite terms of the order $m_c^2/m_{\rm SUSY}^2$:
\begin{equation}
\Sigma _{fi}^{\tilde g LR} \equiv \Sigma _{fi}^{\tilde g LR}\left( 0
\right),\;\;\;\Sigma _{fi}^{\tilde g RL} \equiv \Sigma _{fi}^{\tilde g
  RL}\left( 0 \right),\;\;\;\Sigma _{fi}^{\tilde g LL} \equiv \Sigma
_{fi}^{\tilde g LL}\left( 0 \right),\;\;\;\Sigma _{fi}^{\tilde g RR} \equiv
\Sigma _{fi}^{\tilde g RR}\left( 0 \right)\,.
\label{Sigmaabbrev}
\end{equation}
With these notations the relation between the Yukawa couplings of the MSSM superpotential and the running quark masses of the SM (evaluated at the scale $m_{\rm SUSY}$) is given by
\begin{equation}
{\left[ {{m_{{u_i}}}\left( {1 - \frac{1}{2}\left( {\Sigma _{ii}^{\tilde g LL}
              + \Sigma _{ii}^{\tilde g RR}} \right)} \right) - \Sigma
      _{ii}^{\tilde g LR}} \right]_{{\rm{finite}}}} = {v_u}{Y^{{u_i}}} \, .
\end{equation}
 
\subsection{Renormalization of the squark sector}

\begin{figure*}[t]
\centering
\includegraphics[width=1\textwidth]{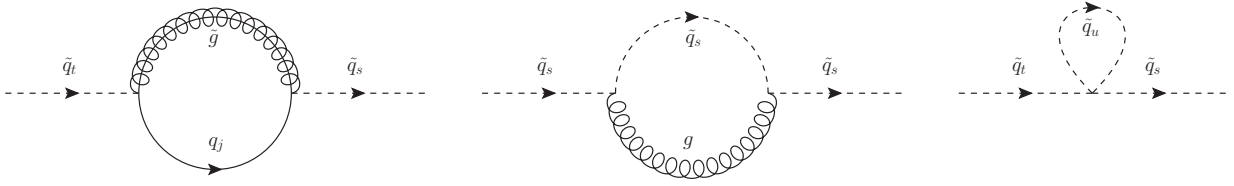}
\caption{Squark self-energy diagrams with SQCD loops: Gluon, gluino and tad-pole contribution (from left to right).}
\label{SquarkSE}
\end{figure*}

As for the quarks, we compute in our approach only LSZ factors corresponding
to flavour-diagonal squark self-energies (i.e. $\tilde u_s \to \tilde u_s$
transitions), while all the other contributions from squark-self energies are
calculated as one-particle irreducible diagrams. The LSZ factors for the squarks then read
\begin{align}
\delta \tilde Z_{\tilde u} ^a &= \dfrac{{{\alpha _s}}}{{4\pi }}{C_F}{\mkern 1mu} (2 + (1 - \xi ))\left[ {\dfrac{1}{\varepsilon } - \dfrac{1}{{{\varepsilon_{\rm{IR}}}}}} \right]{\mkern 1mu} \,,\label{LSZsquarkgluon}\\
\delta \tilde Z_{{\tilde u_s}}^b &= \left.\dfrac{{\partial \Sigma_{\tilde
        u_s\tilde u_s}^{\tilde g + \tilde q}\left( {{p^2}} \right)}}{{\partial {p^2}}}\right|_{p^2=m_{\tilde u_s}^2} =  - \dfrac{{{\alpha _s}}}{{2\pi }}{C_F}\dfrac{1}{\varepsilon } + {\rm{finite}}\,. \label{LSZsquarkgluino}
\end{align}
Like in the quark case $a$ refers to the gluon part and $b$ to the gluino and
squark-tadpole part and $\Sigma _{\tilde u_s\tilde u_s}^{\tilde g + \tilde q}\left(
  {{p^2}}\right)$ denotes the sum of eqs. (\ref{squark_gluino_SE}) and (\ref{squark_squark_SE}). 
From the eqs. (\ref{squark_gluino_SE})-(\ref{squark_squark_SE}) in the
appendix we find that the sum of those UV divergent parts 
of the squark self-energies which are independent of the external momentum
(i.e. the mass-like contribution) is given by
\begin{equation}
\begin{array}{l}
\Sigma _{{{\tilde u}_s}{{\tilde u}_t}}^{{\rm{UVdiv}}}(0) = \dfrac{{{\alpha _s}}}{{2\pi }}{C_F}\dfrac{1}{\varepsilon }\left[ \left(\dfrac{\xi-1}{2}m^2_{\tilde u_s}+m_{\tilde g}^2\right){\delta _{st}} + 2\sum\limits_{j = 1}^3 {\left( {W_{js}^{\tilde u*}m_{{u_j}}^2W_{jt}^{\tilde u} + W_{j + 3,s}^{\tilde u*}m_{{u_j}}^2W_{j + 3,t}^{\tilde u}} \right)}  \right.\\
\;\;\;\;\;\;\;\;\;\;\;\;\;\;\;\;\;\;\;\;\;\;\;\;\;\;\;\;\;\;\;\;\;\; + \sum\limits_{i,j = 1}^3 {\left( {W_{i + 3s}^{\tilde u \star }\Delta _{ij}^{uRL}W_{jt}^{\tilde u} + W_{is}^{\tilde u \star }\Delta _{ij}^{uLR}W_{j + 3t}^{\tilde u}} \right)}  \\
\left. 
\;\;\;\;\;\;\;\;\;\;\;\;\;\;\;\;\;\;\;\;\;\;\;\;\;\;\;\;\;\;\;\;\;\;- 2{m_{\tilde g}}\sum\limits_{j = 1}^3 {\left( {W_{js}^{\tilde u*}{m_{{u_j}}}W_{j + 3,t}^{\tilde u} + W_{j + 3,s}^{\tilde u*}{m_{{u_j}}}W_{jt}^{\tilde u}} \right)}  \right]\,.
\end{array}
\label{SquarkSEdiv}
\end{equation}
To \eq{SquarkSEdiv}, the divergent squark mass terms induced by the LSZ
factors in eqs. (\ref{LSZsquarkgluon}) and (\ref{LSZsquarkgluino})
\begin{equation}
\dfrac{{{\alpha _s}}}{{4\pi }}{C_F} (1 - \xi )\dfrac{1}{\varepsilon } m^2_{\tilde u_s}\delta_{st}\,,\label{deltaMLSZ}
\end{equation}
have to be added, canceling the divergence involving $m^2_{\tilde u_s}$. In order to see the cancellation of the remaining UV divergences in \eq{SquarkSEdiv}, we consider the bare mass matrix which is given in the super-CKM basis
\begin{equation}
{\cal M}_{\tilde u}^{2\left( 0 \right)}\! = \!\left( \begin{array}{*{20}{c}}
{{m}_{U}^{LL2}+\delta{m}_{U}^{LL2} + v_u^2\left( {{{Y}^u}\delta {Y^{u\dag }} + \delta {Y^u}{Y^{u\dag }}} \right)}&{ - {v_u}\left( {{A^u} + \delta {A^u} + \mu \left( {{Y^u} + \delta {Y^u}} \right)\cot \beta } \right)}\\
{ - {v_u}\left( {{A^{u\dag }} + \delta {A^{u\dag }} + \mu \left( {{Y^{u\dag }} + \delta {Y^{u\dag }}} \right)\cot \beta } \right)}& m_{U}^{RR2}+\delta m_{U}^{RR2} + v_u^2\left( {{Y^u}\delta {Y^{u\dag }} + \delta {Y^u}{Y^{u\dag }}} \right)
\end{array} \right)
\end{equation}
Since the squark mixing matrix $W^{\tilde u}$ diagonalizes the renormalized mass matrix, the bare mass matrix is not diagonal in this basis but rather has the form
\begin{equation}
\renewcommand{\arraystretch}{1.8}
\begin{array}{l}
W_{s's}^{\tilde u*}{\left( {{\cal M}_{\tilde u}^{2\left( 0 \right)}}
  \right)_{s't'}}W^{\tilde u}_{t't} = \\
\qquad\;\;\; m_{{{\tilde u}_s}}^2{\delta _{st}} + \sum\limits_{i,j = 1}^3 \left[ v_u^2\left( W_{is}^{\tilde u*}\left( {{Y^u}\delta {Y^{u\dag }} + \delta {Y^u}{Y^{u\dag }}} \right)_{ij}W_{jt}^{\tilde u}   \right.\right.
\\\left.\left.
\qquad + W_{i + 3,s}^{\tilde u*}{\left( {{Y^u}\delta {Y^{u\dag }} + \delta {Y^u}{Y^{u\dag }}} \right)}_{ij}W_{j + 3,t}^{\tilde u} \right)+ W_{is}^{\tilde u*}\delta m^{LL2}_{Uij}W_{jt}^{\tilde u}+ W_{i + 3,s}^{\tilde u*} \delta m^{RR2}_{Uij} W_{j + 3,t}^{\tilde u} \right) \\
\qquad - \left. {v_u}\left( {W_{i + 3s}^{\tilde u \star }{{\left( {\delta {A^{u\dag }} + \mu \delta {Y^{u\dag }}\cot \beta } \right)}_{ij}}W_{jt}^{\tilde u} + W_{is}^{\tilde u \star }{{\left( {\delta {A^u} + \mu \delta {Y^u}\cot \beta } \right)}_{ij}}W_{j + 3t}^{\tilde u}} \right) \right]\,.
\end{array}
\label{deltaMphys}
\end{equation}
Comparing \eq{SquarkSEdiv} and \eq{deltaMLSZ} to \eq{deltaMphys}, we observe that the counterterms 
\begin{equation}
{v_u}\delta {Y^{u_i}} =  - \dfrac{{{\alpha _s}}}{{2\pi }}\dfrac{1}{\varepsilon }{C_F}{m_{u_i}}{\mkern 1mu}\,, \label{deltaYsquark}
\end{equation}
and
\begin{align}
\delta {A}^u_{ij} &= - \dfrac{{{\alpha _s}}}{{2\pi }}\dfrac{1}{\varepsilon }{C_F}\left({A^u_{ij}+2m_{\tilde g}Y^{u_i}\delta_{ij}}\right)\,,\\
(\delta {m}_U^{LL2})_{ij} &= (\delta {m}_U^{RR2})_{ij} =- \dfrac{{{\alpha _s}}}{{2\pi }}\dfrac{1}{\varepsilon }{C_F}m_{\tilde g}^2\delta_{ij}\,.
\end{align}
cancel the divergences. As required by supersymmetry, \eq{deltaYsquark} equals
\eq{deltaYquark}. Therefore, no renormalization of the squark mixing matrices
$W$ is necessary in this formalism\footnote{Furthermore, note that since the
  renormalization of the Yukawa couplings is fixed from the quark sector to be
  in a minimal renormalization scheme, it would not be consistent to absorb
  the finite pieces of the loop-corrections into a redefinition of the squark
  mixing matrices.}. 
	\medskip

In the numerical analysis, we will use the connection between the on-shell and
the $\overline{\rm DR}$ mass. This relation is given by
\begin{equation}
m_{{\tilde u_s}}^{2\;{\rm{OS}}} = m_{{\tilde u_s}}^{2\;\overline {{\rm{DR}}} }
+ \Sigma _{{\tilde u_s}{\tilde u_s}}^{\rm finite}\left( {{p^2} = m_{{\tilde u_s}}^2} \right)\,.
\label{shiftdrbaros}
\end{equation}
\medskip

\subsection{Gluon contributions}

\begin{figure*}[t]
\centering
\includegraphics[width=0.7\textwidth]{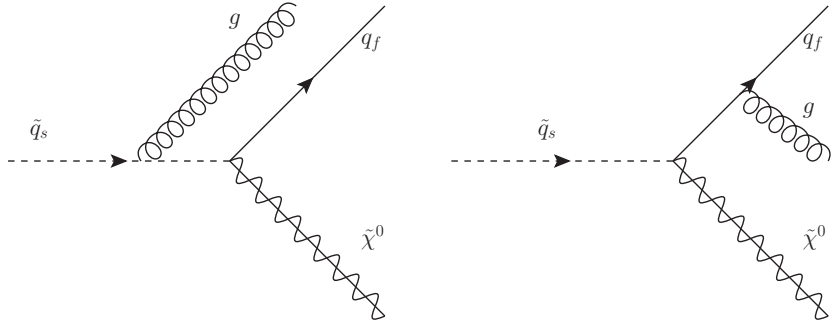}
\caption{Feynman diagrams showing the real emission of a gluon, i.e. the process $\tilde u_1\to c+g+\tilde\chi^0_1$.}
\label{RealEmission}
\end{figure*}

Here we combine the virtual gluon contributions with the real radiation (see
Fig~\ref{RealEmission}) and show the cancellation of the infrared and
collinear divergences. In our calculations all singularities are regularized 
dimensionally; more precisely, we use dimensional reduction
and introduce the renormalization scale in the form $\mu^2
e^{\gamma}/(4\pi)$, where $\gamma=0.577...$ is the Euler constant.  
\medskip
 
\begin{figure*}[t]
\centering
\includegraphics[width=0.7\textwidth]{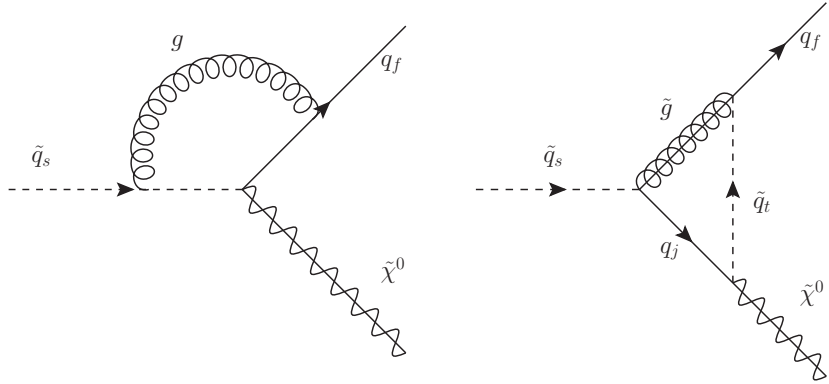}
\caption{Genuine vertex corrections involving gluons (left diagram) and gluinos (right diagram).}
\label{Genuine_Vertex_Corrections}
\end{figure*}

For the vertex correction diagram due to gluon exchange (left diagram in
Fig.~\ref{Genuine_Vertex_Corrections}) we get
\begin{eqnarray}
\label{vertex}
V^g= & {\cal A}_0 \dfrac{\alpha_s}{4 \pi} \, C_F \left[
  \dfrac{(1-(1-\xi))}{\varepsilon}-\dfrac{1}{\varepsilon_{\rm{IR}}^2} +
\dfrac{-2+(1-\xi) - 2 L_{\mu} + 2 \ln(1-x_1)}{\varepsilon_{\rm{IR}}} -2 - \dfrac{\pi^2}{12}
\right. \nonumber \\
& \left.  -2 L_{\mu}^2 - 2 L_{\mu} + 4 L_{\mu} \ln(1-x_1) - 2
\ln^2(1-x_1) + 2 \ln(1-x_1) - 2 {\rm Li}_2(x_1) \, \right] \, ,
\end{eqnarray}
using the abbreviations $x_1=m_{{\tilde\chi^0_1}}^2/m_{\tilde u_1}^2$ and
$L_{\mu}= \ln(\mu/m_{\tilde u_1})$. $\xi$ denotes the gauge-parameter
which is involved in the gluon propagator. As before, poles of the form $1/\varepsilon$
correspond to ultraviolet singularities, while poles of the form
$1/\varepsilon^2_{\rm{IR}}$, $1/\varepsilon_{\rm{IR}}$ are due to infrared and collinear
singularites. Finally ${\cal A}_0$
is the tree-level amplitude (originating from
\eq{squark-quark-neutralino-vertex}), reading 
\begin{equation}
{\cal A}_0 =
i \bar{u}(p_{u_2}) \left( \Gamma_{\tilde u_1 u_2}^{\tilde{\chi}_1^0 L \star} P_R
 + \Gamma_{\tilde u_1 u_2}^{\tilde{\chi}_1^0 R \star} P_L
 \right) v(p_{\tilde{\chi}_1^0})\,.
\end{equation}

To get the renormalized result ${\cal A}^g$ for the amplitude, we need to add the contributions
induced by the gluon part of the LSZ factors of
the (massless) charm quark and the stop squark (see \eq{LSZquarkgluon} and
\eq{LSZsquarkgluon}, respectively), as well as the effects induced by the
renormalization
constants for the coupling constants $e$ and $Y^{u_i}$ appearing in the tree-level
squark-quark-neutralino vertex. These renormalization
constants are written as $Z_e=1 + \delta Z_e^a  + \delta Z_e^b$ for the gauge
coupling $e$ 
and $Z_{Y^{u_i}} = 1+ \delta Z_{Y^{u_i}}^a + \delta Z_{Y^{u_i}}^b $ for the Yukawa coupling $Y^{u_i}$. The parts due gluon
corrections, which are relevant in this subsection, read
\footnote{The parts  $\delta Z_e^b$ and  $\delta Z_{Y^{u_i}}^b$, which
are due to gluino corrections, will be taken into account in the following subsection.}
\begin{equation}
\label{couplingrenorm}
\delta Z_e^a = \delta Z_{Y^{u_i}}^a =
- \dfrac{\alpha_s}{4 \pi} C_F \, \dfrac{3}{2} \, \dfrac{1}{\varepsilon} \, .
\end{equation}
As expected, these expressions are independent of the gauge parameter $\xi$.
Adding up the mentioned contributions, we get the renormalized amplitude
\begin{eqnarray}
{\cal A}^g = && {\cal A}_0 \dfrac{\alpha_s}{4 \pi} \, C_F \, \left[
-\dfrac{1}{\varepsilon_{\rm{IR}}^2}
+ \dfrac{-5/2 - 2 L_{\mu} + 2 \ln(1-x_1)}{ \varepsilon_{\rm{IR}}} -2 - \dfrac{\pi^2}{12}
\right. \nonumber \\
&& \left. -2 L_{\mu}^2 - 2 L_{\mu} + 4 L_{\mu} \ln(1-x_1) - 2
\ln^2(1-x_1) + 2 \ln(1-x_1) - 2 {\rm Li}_2(x_1) \right] \, .
\end{eqnarray}
This result is, as required by consistency, again independent of the gauge parameter $\xi$.
To get from the renormalized amplitude to the decay width is
straightforward. Doing all these manipulations in $d=4-2\varepsilon$ dimensions, we get
\begin{eqnarray}
\label{virt}
\Gamma^{{\rm virt}} = && \Gamma_0 \dfrac{\alpha_s}{4 \pi} \, C_F \, \left[
-\dfrac{2}{\varepsilon_{\rm{IR}}^2}
+ \dfrac{-9 - 8 L_{\mu} + 8 \ln(1-x_1)}{ \varepsilon_{\rm{IR}}} -22 + \dfrac{\pi^2}{3}
 -16 L_{\mu}^2 - 30 L_{\mu} \right. \\
&& \qquad\qquad\left. + 32 L_{\mu} \ln(1-x_1) - 16
\ln^2(1-x_1) + 30 \ln(1-x_1) - 4  {\rm Li}_2(x_1) \, 
\right] \, ,\nonumber 
\end{eqnarray}
where $\Gamma_0$ is the corresponding decay width at order $\alpha_s^0$ given in \eq{tree-level-decay-width}.
\medskip

We now turn to the bremsstrahlung corrections (see Fig~\ref{RealEmission}). 
Using the information given in section~\ref{sec:phasespacemassive} of the appendix on the
three-particle phase space and making use of the mathematica package HypExp 2.0 \cite{Huber:2005yg}, it is straightforward to
derive the decay width for $\tilde{u}_1 \to c \tilde{\chi}_1^0 g$. 
We obtain 
\begin{eqnarray}
\label{brems}
\Gamma^{{\rm brems}} = && \Gamma_0 \dfrac{\alpha_s}{4 \pi} \, C_F \, \left[
\dfrac{2}{\varepsilon_{\rm{IR}}^2}
+ \dfrac{9 + 8 L_{\mu} - 8 \ln(1-x_1)}{ \varepsilon_{\rm{IR}}} - \dfrac{5 \pi^2}{3}
+\dfrac{69-71x_1}{2(1-x_1)} + 16 L_{\mu}^2 + 36 L_{\mu}
\right. \nonumber \\
&&  - 32 L_{\mu} \ln(1-x_1) + 16
\ln^2(1-x_1)  - 4 \, (9+\ln(x_1)) \, \ln(1-x_1)  \nonumber \\
&& \left.  - \dfrac{x_1(4-3x_1)}{(1-x_1)^2} \ln(x_1)
- 4 {\rm Li}_2(x_1) \, 
\right] \, .
\end{eqnarray}
Adding the virtual corrections (\ref{virt}) and the gluon bremsstrahlung
corrections (\ref{brems}), we get
\begin{eqnarray}
\label{virtbrems}
\Gamma^{{g}} = && \Gamma_0 \dfrac{\alpha_s}{4 \pi} \, C_F \, \left[ -
\dfrac{4 \pi^2}{3} +\dfrac{25-27x_1}{2(1-x_1)}  + 6 L_{\mu}
  - 2 \, (3 + 2 \ln(x_1)) \, \ln(1-x_1) \right. \nonumber \\
&& \qquad\left.  - \dfrac{x_1(4-3x_1)}{(1-x_1)^2} \ln(x_1)
- 8 {\rm Li}_2(x_1) \, 
\right] \, .
\end{eqnarray}
As expected, the collinear and infrared singularities canceled and the result is finite\footnote{As the renormalization scheme in Ref.~\cite{Grober:2014aha} is quite different from ours, a full comparison is difficult.
It was, however, possible to compare the gluino vertex correction, the virtual gluon
corrections and the gluon bremsstrahlung corrections individually. Taking into account
that in Ref.~\cite{Grober:2014aha}  the two-particle phase space (and a corresponding
 part of the three-particle phase space) is in $d=4$ dimensions and that
the renormalization scale is of the form
 $\mu^{2\varepsilon} \Gamma(1-\varepsilon)/(4\pi)^{\varepsilon}$),
we found that the results are in agreement. In our calculation
we used a $d$-dimensional phase space (and introduce the
renormalization scale in the form $\mu^{2\varepsilon}
e^{\gamma \varepsilon}/(4\pi)^{\varepsilon}$).}.
\medskip

\subsection{Gluino and squark-tadpole contributions}

We write the amplitude containing the tree-level and the contribution of loop
diagrams involving gluinos and the squark tadpole (right diagram in Fig~\ref{SquarkSE}) as 
\begin{equation}
{\cal A}^{\tilde g}=i\bar u(p_{u_2})\left[ {\left( {\Gamma _{{{\tilde u}_1}{u_2}}^{\tilde \chi _1^0L*} + 
\Lambda _{{{\tilde u}_1}{u_2}}^{\tilde \chi _1^0L*}} + 
\sum\limits_{j = 1}^3 {X_{u_j u_2}^{L*}\Gamma _{{{\tilde u}_1}{u_j}}^{\tilde \chi
    _a^0L*}}  + \sum\limits_{s = 1}^6
 {\Gamma _{{{\tilde u}_s}{u_2}}^{\tilde \chi _1^0L*}\tilde X_{\tilde u_s
     \tilde u_1}} \right){P_R}   + 
(R \leftrightarrow L)} \right] v(p_{\tilde \chi^0_1})\,.
\label{gluino-contribution}
\end{equation}
Here $\Gamma _{{{\tilde u}_1}{u_2}}^{\tilde \chi _1^0L*}$ encodes the tree-level contribution and
$\Lambda_{{{\tilde u}_1}{u_2}}^{\tilde \chi _1^0*}$, given in Eq. (\ref{GluinoVertexCorrection}) of
the appendix, denotes the genuine vertex correction involving the gluino.
Furthermore, $X^{L,R}_{u_f u_i}$ and $\tilde X_{\tilde u_s \tilde u_t}$ originate from quark and squark self-energy diagrams, respectively. The explicit expressions read
\begin{align}
\renewcommand{\arraystretch}{2.2}
\tilde X_{\tilde u_s \tilde u_t} &= \left\{ {\begin{array}{*{20}{c}}
{\dfrac{\Sigma_{\tilde u_s\tilde u_t}(p^2=m_{\tilde u_t}^2)}{{m_{{{\tilde u}_t}}^2 - m_{{{\tilde u}_s}}^2}}\;\;{\rm{for}}\;\;s \ne t}\,,\\
{\dfrac{1}{2}\delta \tilde Z^b_{\tilde u_s}\;\;\;{\rm{for}}\;\;s = t}\,,
\label{tildeX}
\end{array}} \right. \\
X_{u_f u_i}^L &= \left\{ {\begin{array}{*{20}{c}}
{\dfrac{{{m_i}\Sigma _{fi}^{\tilde g LR} + {m_f}\Sigma _{fi}^{\tilde g RL} + m_i^2\Sigma _{fi}^{\tilde g LL} + {m_f}{m_i}\Sigma _{fi}^{\tilde g RR}}}{{m_i^2 - m_f^2}}\;\;{\rm{for}}\;\;f \ne i}\,,\\
{\dfrac{1}{2}\delta
  Z^{Lb}_{u_f}\;\;\;\;\;\;\;\;\;\;\;\;\;\;\;\;\;\;\;\;\;\;\;\;\;\;\;\;\;\;\;\;\;\;\;\;\;\;\;\;\;\;\;\;\;{\rm{for}}\;\;f
  = i} \, .
\end{array}} \right.
\end{align}
\medskip

Let us briefly discuss the ultraviolet singularities in
Eq. (\ref{gluino-contribution}) and how they get canceled: 
All divergences in the off-diagonal elements of $\tilde X_{\tilde u_s \tilde u_t}$ are canceled by the
counter-terms induced through the renormalization of $Y^{u_i}$, $A^u_{ij}$,
$\left( m_U^{LL2}\right)_{ij}$ and  $\left( m_U^{RR2}\right)_{ij}$
 in the squark mass matrix, while the
off-diagonal elements of $X_{u_f u_i}^{L,R}$ are finite ab initio.
Therefore, we are effectively left in Eq. (\ref{gluino-contribution}) with the
singularities in the flavour conserving parts of $X^{L,R}_{u_f u_i}$ and
$\tilde X_{\tilde u_s \tilde u_t}$ which originate from LSZ factors, and with
the singularities present in the vertex correction 
$\Lambda _{{{\tilde u}_1}{u_2}}^{\tilde \chi _1^0L*}$. Using 
the unitarity of the squark-mixing matrices in \eq{GluinoVertexCorrection},
the latter singularities read 
\begin{align}
\Lambda _{\tilde{u}_1 u_2;\rm{div}}^{\tilde \chi _1^0L*} &= \frac{{ - {\alpha _s}}}{{2\pi }}{C_F}\frac{1}{\varepsilon }W_{2+3,1}^{\tilde u}{Y^{{u_2}}}Z_N^{41*} \,,\nonumber\\
\Lambda _{\tilde{u}_1 u_2;\rm{div}}^{\tilde \chi _1^0 R*} &= \frac{{ -{\alpha _s}}}{{2\pi }}{C_F}\frac{1}{\varepsilon }W_{2,1}^{\tilde u}{Y^{u_2*}}Z_N^{41} \,.\label{GluinoVertexDiv}
\end{align}
It is straightforward to see that the remaining singularities get cancelled
against those which are induced by the gluino parts $\delta Z_e^b$ and $\delta
Z_{Y^{u_i}}^b$ of the renormalization constants of the gauge coupling $e$ and 
$Y^{u_i}$ present in the tree-level squark-quark-neutralino vertex. These
renormalization constants read\footnote{$\delta Z_e^b=-\delta{Z_e^b}$ verifies
  that the electric charge is not renormalized by SQCD
and the compatibility of $\delta Z_{Y^{u_i}}^a +\delta Z_{Y^{u_i}}^b$ with \eq{deltaYquark} shows
that SUSY is respected.}
\begin{align}
\label{couplingrenormadd}
\delta Z_e^b &=-\delta Z_e^a \,,\\
\delta Z_{Y^{u_i}}^b &=- \dfrac{\alpha_s}{8 \pi} C_F \, \dfrac{1}{\varepsilon}\,.
\end{align}
where $Z_e^a$ is given in \eq{couplingrenorm}.
\medskip

Therefore, the renormalized version of the amplitude is obtained by just
taking the finite part of \eq{gluino-contribution}. The corresponding
contribution to the decay width is then obtained by inserting the renormalized
amplitude into \eq{tree-level-decay-width} and working out the interference term,
i.e. the term proportional to $\alpha_s$.

\begin{figure*}[t]
\centering
\includegraphics[width=0.95\textwidth]{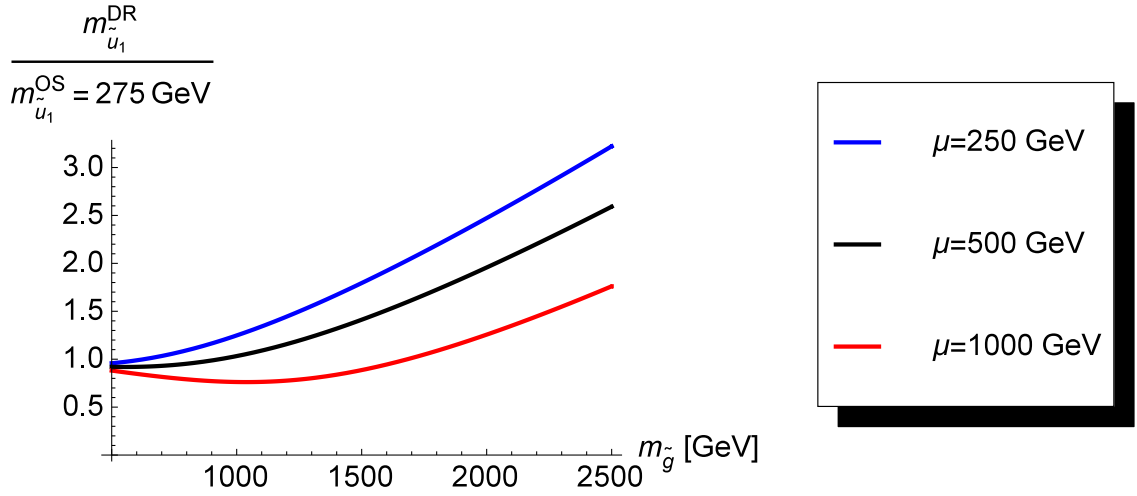}
\caption{Ratio of the $\overline{\rm DR}$ stop mass over its on-shell mass for
  $m^{\rm OS}_{\tilde u_1}= 275$~GeV as generated by \eq{massmatu2} and
  $A^t=1$ TeV as a function of the gluino mass for different values of the
  renormalization scale $\mu$ (see text). }
\label{Shift-OnShell-DR}
\end{figure*}

\section{Numerical Analysis}\label{numerics}

In our numerical analysis we investigate the size of the calculated SQCD
corrections. For this purpose we consider the following squark mass matrix given in the $\overline{\rm DR}$-scheme:
\begin{equation}
{\cal M}_{\tilde u}^2 = \left( \begin{array}{*{20}{c}}
\left( {2\;{\rm{TeV}}} \right)^2&0&0&0&0&0\\
0&{{{\left( {2\;{\rm{TeV}}} \right)}^2}}&{\Delta _{23}^{LL}}&0&0&{\Delta _{23}^{LR}}\\
0&{\Delta _{23}^{LL*}}&{{\left( m^{LL}_{33} \right)^2}}&0&{\Delta _{23}^{RL*}}&{ - {v_u}A^t}\\
0&0&0&{{{\left( {2\;{\rm{TeV}}} \right)}^2}}&0&0\\
0&0&{\Delta _{23}^{RL}}&0&{{{\left( {2\;{\rm{TeV}}} \right)}^2}}&{\Delta _{23}^{RR}}\\
0&{\Delta _{23}^{LR*}}&{ - {v_u}A^t}&0&{\Delta _{23}^{RR*}}&{{\left( m^{RR}_{33} \right)}^2}
\end{array} \right)\,.
\label{massmatu2}
\end{equation}
Here $\Delta _{ij} = \delta_{ij} \sqrt{{\cal M}_{\tilde u,ii}^2 {\cal M}_{\tilde u,jj}^2}$ 
parametrizes the flavour change (and is assumed to be small compared to the diagonal elements) and we choose $A^t=\pm 1$ TeV. In the following, we will consider the case of $m^{RR}_{33}=m^{LL}_{33}$ (i.e. maximal mixing). For the neutralino, which we assume to be bino like, we choose a mass of 250~GeV and use $\alpha_s(m_{\rm SUSY})=0.087$ as an input.
\medskip

At tree-level, the scheme for the stop mass is not defined. At the 1-Loop level
 the quantities of the MSSM superpotential must be renormalized in a process independent way in order to respect
supersymmetry, e.g. the Yukawa couplings have to be renormalized in the
$\overline{\rm DR}$-scheme. For consistency, also all other elements of 
the squark mass matrix should be renormalized in this scheme 
as well and should be given at the same renormalization scale. After
diagonalization of the squark mass matrix, the eigenvalues 
correspond to $\overline{\rm DR}$-masses which can be
translated to on-shell masses if necessary or desired. This is the case for $\tilde u_1\to c\tilde\chi^0_1$ 
where the masses entering the decay width in \eq{tree-level-decay-width} are on-shell
masses. 
\medskip

The shift between the $\overline{\rm DR}$ and the on-shell mass 
(see Eq. \ref{shiftdrbaros}) turns out to be numerically especially important for our scenario with a light stop because
it scales like \footnote{Even though the correction is very large, pertubation
  theory still works, because the parametric enhancement $m_{\tilde g}^2/m_{\tilde q}^2$
can only appear once at any loop-level. Therefore, higher loop-corrections
will have the size of ordinary SQCD effects compared to the one-loop result.}
$m_{\tilde g}^2/m_{\tilde q}^2$. In Fig.~\ref{Shift-OnShell-DR} we show the
ratio $m^{\overline{\rm DR}}_{\tilde u_1}/m^{\rm OS}_{\tilde u_1}$ as
a function of the gluino mass at the 1-loop level for $m_{\tilde u_1}^{\rm OS}=275$~GeV. 
For this we set all flavour off-diagonal elements $\Delta_{ij}$ in \eq{massmatu2} to zero.
\medskip

Note that for large gluino masses the on-shell stop mass is smaller than the
$\overline{\rm DR}$ mass. This has interesting consequences for model building
with light stops: Assuming that there is already a splitting between the
$\overline{\rm DR}$ squark masses of the first two generations and the stop squark (for example
due to the running from the GUT scale to the SUSY scale) at the SUSY scale,
then this splitting is significantly increased for heavy gluinos, making the
stop even lighter. Therefore, light stop scenarios, which are interesting for
the decay $\tilde u_1\to c\tilde \chi^0_1$, can be even generated via finite
loop effects.
\medskip

For the numerical analysis of the SQCD corrections to $\tilde u_1\to
  c\tilde \chi^0_1$, we choose $m_{33}^{LL}=m_{33}^{RR}$ in
  Eq. (\ref{massmatu2}) in such a way, that
  a given on-shell mass for $\tilde u_1$ (275 GeV in our example) results
  after diagonalizing Eq.~(\ref{massmatu2}) and shifting the so-obtained
  $\overline{\rm DR}$ squark masses to the corresponding on-shell
  masses \footnote{$m_{33}^{LL}=m_{33}^{RR}$ determined in this way will depend
  on the gluino mass $m_{\tilde g}$ and on the renormalization scale $\mu$.}. This
  procedure we do for both, the tree-level decay width $\Gamma^{\rm tree}$ and
  for the SQCD corrected version $\Gamma^{\rm 1-loop}$ calculated in this paper.
\medskip

\begin{figure*}[t!]
\centering
\includegraphics[width=0.8\textwidth]{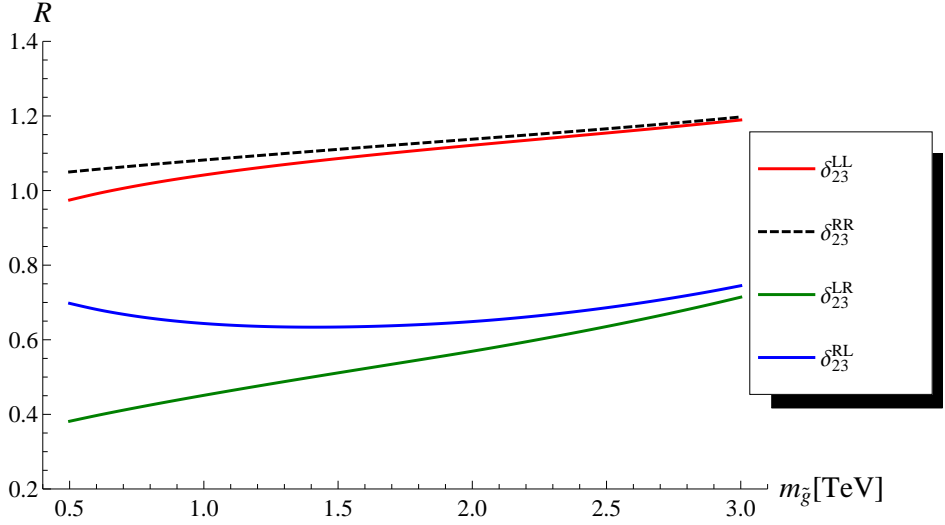}
\caption{Ratio of the decay width including the 1-loop SQCD corrections over
  the tree-level decay width for different sources of flavour violation as a
  function of the gluino mass for $A^t=1$ TeV. The
  renormalization scale is chosen to be $\mu=275$ GeV.}
\label{1loopOVERtree1}
\end{figure*}

\begin{figure*}[t!]
\centering
\includegraphics[width=0.8\textwidth]{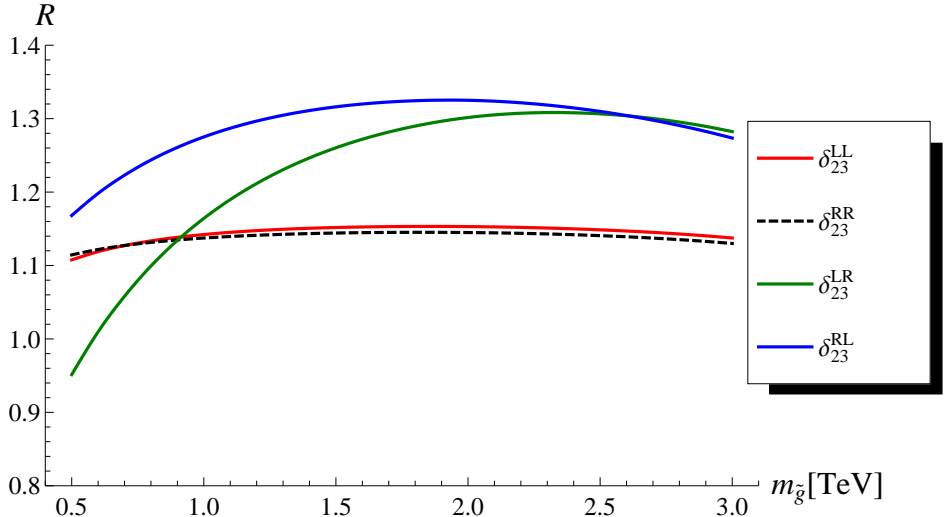}
\caption{Ratio of the decay width including the 1-loop SQCD corrections over
  the tree-level decay width for different sources of flavour violation as a
  function of the gluino mass for $A^t=-1$ TeV. The renormalization scale is chosen to be $\mu=275$ GeV.}
\label{1loopOVERtree2}
\end{figure*}

In Fig.~\ref{1loopOVERtree1} (Fig.~\ref{1loopOVERtree2}) we illustrate the
effect of the one-loop contributions for positive (negative) $A^t$ for the
four different sources of flavour-violation: $\delta _{23}^{RR}$, $\delta
_{23}^{LL}$, $\delta _{23}^{RL}$ and $\delta _{23}^{LR}$. Here we defined the
ratio $R=\Gamma^{\rm 1-loop}/\Gamma^{\rm tree}$ of the partial widths. In each
of the four curves in Fig.~\ref{1loopOVERtree1} and Fig.~\ref{1loopOVERtree2}
the indicated $\delta_{ij}^{AB}$ is put to 0.01, while
the other $\delta$'s are switched off. Note that the actual numerical values of the mentioned $\delta$s
drops out in this ratio to a very good approximation. We find that if bilinear terms
are the only sources of flavour violation, the SQCD
effects are around $10\%$, while if flavour violations originate from
trilinear terms the corrections can reach $\pm 50\%$ or even more. The large
corrections in the case of $\delta _{23}^{RL}$ and $\delta _{23}^{LR}$ can be
traced back to the suppressed decay width for left-handed charm quarks.
\medskip

\section{Conclusions}\label{conclusions}

In this article we computed the 1-loop SQCD corrections to the decay $\tilde u_1\to c\tilde \chi^0_1$ in the MSSM with generic sources of flavour violation. This decay is phenomenologically very important if the mass splitting between the neutralino and the lightest stop is smaller than the top mass. In particular, we pointed out that a sizable partial width for $\tilde u_1\to c\tilde \chi^0_1$, which is possible in the presence of non-minimal sources of flavour violation, can significantly weaken the LHC exclusion bounds obtained from $\tilde u_1 \to W b \chi^0_1$ where usually a branching ratio of $100\%$ is assumed.
\medskip

Working in the super-CKM basis with diagonal Yukawa couplings and renormalizing all parameters in the $\overline{\rm DR}$ scheme, we explicitly checked for the cancellation of UV divergences and verified that SUSY relations are satisfied. In particular, in the squark sector all divergences are eliminated by flavour-conserving counter-terms to Yukawa couplings, $A$-terms and the bilinear terms, meaning that no renormalization of the squark mixing matrices is necessary. Concerning the gluon corrections we regularized all divergences dimensionally and verified their cancellation in a general $R_\xi$ gauge.
\medskip

Numerically, we observe a large shift between the on-shell and the
$\overline{\rm DR}$ mass of the stop. Due to the inherited quadratic
divergence, the shift involves a term proportional to $m_{\tilde
  g}^2/m_{\tilde q}^2$. Since for large gluino masses the on-shell stop mass
is driven to smaller values compared to the $\overline{\rm DR}$ mass, it is
important to take into account this shift for model building. Taking the
on-shell stop mass as in input, we find a SQCD enhancement of the decay
width compared to the tree-level for $\tilde u_1\to c\tilde \chi^0_1$
(assuming a bino like LSP) of approximately $10\%$ if the flavour violation is
due to bilinear terms and $\pm 50\%$ and more if the single origin of flavour
violation are the trilinear terms.
\medskip

For the future, a NLO SQCD calculation of $\tilde u_1 \to W b \chi^0_1$ would be desirable and a phenomenological study of the impact of $\tilde u_1\to c\tilde \chi^0_1$ on the exclusion bounds from $\tilde u_1 \to W b \chi^0_1$ is planed.
\medskip

\begin{acknowledgments}
This work is supported by the Swiss National Science Foundation. A.C. is supported by a Marie Curie Intra-European Fellowship of the European Community's 7th Framework Programme under contract number (PIEF-GA-2012-326948). We like to thank the authors of Ref.~\cite{Grober:2014aha} for their kind help in comparing
the details of the real emission part and for useful comments on the
manuscript.
\end{acknowledgments}

\vspace{1cm}

\appendix

{\bf Appendix}

\subsection{Relevant phase-space formulas}
\label{sec:phasespacemassive}
The fully differential decay width $d\Gamma$ for a generic process 
$p \to p_1 + p_2 + ... + p_n$ can be written as
\begin{equation}
d\Gamma = \dfrac{1}{2m} \, \overline{|M|^2} \, D\Phi(1 \to n) \, ,
\end{equation}
where  $\overline{|M|^2}$ is the squared matrix element, summed and
  averaged
over spins and colors of the particles in the final and initial
state, respectively, and $m$ is the mass of the decaying particle.
\medskip

In ref. \cite{Asatrian:2012tp,Asatrian:2014mwa}
useful parametrizations for the
phase-space factors $D\Phi(1 \to n)$ have been given for $n=3,4$,
for the case where all final-state particles are massive. In our problem
we only use the case $n=3$ where only the neutralino is massive;
this means that the general formula simplifies.
In the following subsection we see that the 3-particle
phase-space can be parametrized in terms of two parameters
$\lambda_1$ and $\lambda_2$, which run independently in the range
$[0,1]$. Of course, all scalar products involved
in  $\overline{|M|^2}$ can be expressed in terms of these parameters.
\subsubsection{Phase-space parametrization for the 3-particle final
  state}
\label{subsec:phasespace3}
In our application we identify $p_1$ with the neutralino, $p_2$ with the
(massless) charm quark and  $p_3$ with the gluon and define
$x_1=m_{\tilde{\chi}_1^0}^2/m_{\tilde u_1}^2$. Starting from
eq. (2.10) of ref. \cite{Asatrian:2012tp}, one gets
\begin{eqnarray}
D\Phi(1\to 3)= && \dfrac{{m}_{\tilde u_1}^{2 d-6}2^{1-2 d} \pi^{1-d}}{\Gamma(d-2)}
 [(1-\lambda_1)
   \lambda_1]^{\dfrac{d-4}{2}} [(1-\lambda_2)
   \lambda_2]^{d-3} \times \nonumber \\
&& (1-{x}_1)^{2d-5} [\lambda_2
   (1-{x}_1)+{x}_1]^{\dfrac{2-d}{2}} \, d\lambda_1 d\lambda_2 \, .  
\end{eqnarray}
The scalar products of the momenta $p_i$, encoded in the quantities
$s_{ij}=(p_i+p_j)^2/m_{\tilde u_1}^2$, can be written in terms of the
parameters $\lambda_1$ and $\lambda_2$ as
\begin{eqnarray}
s_{13}&=&\lambda_2 (1-{x}_1)+{x}_1 \nonumber\\
s_{12}&=&\dfrac{\lambda_1 (\lambda_2-1) \lambda_2
   (1-{x}_1)^2-{x}_1}{\lambda_2
   ({x}_1-1)-{x}_1}\nonumber \, .
\end{eqnarray}
\medskip

\subsection{Loop functions}
The one-loop functions which appear at various places in this appendix are defined as
\begin{equation}
A_0(m^2) = \dfrac{16 \pi^2}{i} \dfrac{\mu^{2\varepsilon}
  e^{\gamma \varepsilon}}{(4\pi)^{\varepsilon}} \, \int \dfrac{d^d
  \ell}{(2\pi)^d} \, \dfrac{1}{[\ell^2-m^2]}
\end{equation}
\begin{equation}
\begin{aligned}
B_0(p^2;m_1^2,m_2^2)& =\dfrac{16 \pi^2}{i} \dfrac{\mu^{2\varepsilon}
  e^{\gamma \varepsilon}}{(4\pi)^{\varepsilon}} \, \int \dfrac{d^d
  \ell}{(2\pi)^d} \, \dfrac{1}{[\ell^2-m_1^2] \, [(\ell+p)^2-m_2^2]} \\
&=  \mu ^{2 \varepsilon }e^{\gamma  \varepsilon } \Gamma (\varepsilon )\int_0^1{\left[-x
\left(m_1^2-m_2^2+p^2\right)+m_1^2+p^2 x^2\right]{}^{-\varepsilon }}
\end{aligned}
\end{equation}
\begin{equation}
\begin{aligned}
B_1(p^2;m_1^2,m_2^2) \, p^{\mu} &= \dfrac{16 \pi^2}{i} \dfrac{\mu^{2\varepsilon}
  e^{\gamma \varepsilon}}{(4\pi)^{\varepsilon}} \, \int \dfrac{d^d
  \ell}{(2\pi)^d} \, \dfrac{\ell^{\mu}}{[\ell^2-m_1^2] \, [(\ell+p)^2-m_2^2]} \\
&= p^{\mu}\frac{A_0(m_1^2)-A_0(m_2^2)-(p^2+m_1^2-m_2^2)B_0(p^2;m_1^2,m_2^2)}{2p^2}
\end{aligned}
\end{equation}
\begin{equation}
\begin{aligned}
B_2(p^2;m_1^2,m_2^2) &= \dfrac{16 \pi^2}{i} \dfrac{\mu^{2\varepsilon}
  e^{\gamma \varepsilon}}{(4\pi)^{\varepsilon}} \, \int \dfrac{d^d
  \ell}{(2\pi)^d} \, \dfrac{\ell^2}{[\ell^2-m_1^2] \, [(\ell+p)^2-m_2^2]} \\
&=A_0(m_2^2)+m_1^2B_0(p^2;m_1^2,m_2^2)
\end{aligned}
\end{equation}
\begin{equation}
 \begin{aligned}
C_0(p_1^2,(p_1-p_2)^2&,p_2^2;m_0^2,m_1^2,m_2^2) \\
&=\dfrac{16 \pi^2}{i}
\dfrac{\mu^{2\varepsilon}
  e^{\gamma \varepsilon}}{(4\pi)^{\varepsilon}} \, \int \dfrac{d^d
  \ell}{(2\pi)^d} \, \dfrac{1}{[\ell^2-m_0^2] \, [(\ell+p_1)^2-m_1^2] \,
[(\ell+p_2)^2-m_2^2]} \\
&=- \mu ^{2 \varepsilon }e^{\gamma  \varepsilon } \Gamma (\varepsilon
+1)\int_0^1dx\int_0^{1-x}{dy}\left[-x \left(m_0^2-m_1^2+p_1^2\right)\right.
\\
&\left.-y \left(m_0^2-m_2^2+p_2^2\right)+m_0^2+p_1^2 x^2+2 x y
p_1\cdot p_2+p_2^2 y^2\right]{}^{-(1+\varepsilon)}
\end{aligned}
\end{equation}

\begin{equation}
 \begin{aligned}
C_2(p_1^2,(p_1-p_2)^2&,p_2^2;m_0^2,m_1^2,m_2^2) \\
&=\dfrac{16 \pi^2}{i}
\dfrac{\mu^{2\varepsilon}
  e^{\gamma \varepsilon}}{(4\pi)^{\varepsilon}} \, \int \dfrac{d^d
  \ell}{(2\pi)^d} \, \dfrac{\ell^2}{[\ell^2-m_0^2] \, [(\ell+p_1)^2-m_1^2] \,
[(\ell+p_2)^2-m_2^2]} \\
&=B_0((p_2-p_1)^2;m_1^2,m_2^2)+m_0^2C_0(p_1^2,(p_1-p_2)^2,p_2^2;m_0^2,m_1^2,m_2^2)
\end{aligned}
\end{equation}

\subsection{Vertex correction involving the gluino}
The correction of the squark-quark-neutralino vertex involving the gluino (see
right frame of Fig. \ref{Genuine_Vertex_Corrections}) reads
\begin{equation}
\begin{aligned}
\Lambda^{\tilde \chi^0_1*}_{\tilde{u}_1
  u_2}=&\dfrac{{-1}}{16{\pi^2}}\sum\limits_{j,s}^{}\left[\Gamma_{\tilde u_s
    u_2}^{\tilde gL*}\left( {{C}_2}\Gamma _{\tilde u_1 u_j}^{\tilde
    gR*}\Gamma _{{{\tilde u}_s}{u_j}}^{\tilde \chi _1^0R} + {C}_0\Gamma
  _{\tilde{u}_1 u_j}^{\tilde gL*}\Gamma _{{{\tilde u}_s}{u_j}}^{\tilde \chi _1^0R}m_{\tilde g}^{}{{m}_{{u_j}}} + \left({{{C}_0} + {C_{{p_{\tilde u_1}}}}} \right)\Gamma _{\tilde u_1{u_j}}^{\tilde gL*}\Gamma _{{{\tilde u}_s}{u_j}}^{\tilde \chi _1^0L}m_{\tilde g}^{}{m_{\tilde \chi_1^0}} \right.\right. \\ 
&\left.\left. + {C_{{p_{\tilde u_1}}}}\Gamma _{\tilde u_1{u_j}}^{\tilde gR*}\left(
      {\Gamma _{{{\tilde u}_s}{u_j}}^{\tilde \chi _1^0R}m_{\tilde u_1}^2 +
        \Gamma _{{{\tilde u}_s}{u_j}}^{\tilde \chi
          _1^0L}{{m}_{{u_j}}}{m_{\tilde \chi _1^0}}} \right)
  \right){P_R}+ (L \leftrightarrow R) \right]
\label{GluinoVertexCorrection}
\end{aligned}
\end{equation}
with the abbreviations:
\begin{align}
 C_0&\equiv C_0(m_{\tilde u_1}^2,m_{\tilde \chi_1^0}^2,0;m_{\tilde g}^2,m_{u_j}^2,m_{\tilde u_s}^2) \\
C_2&\equiv C_2(m_{\tilde u_1}^2,m_{\tilde \chi_1^0}^2,0;m_{\tilde g}^2,m_{u_j}^2,m_{\tilde u_s}^2) \\
C_{p_{\tilde u_1}}&\equiv C_{p_{\tilde u_1}}(m_{\tilde u_1}^2,m_{\tilde
  \chi_1^0}^2,0;m_{\tilde g}^2,m_{u_j}^2,m_{\tilde u_s}^2) \, .
\end{align}
$C_{p_{\tilde u_1}}$ is defined through the decomposition
\begin{equation}
 \begin{aligned}&\dfrac{16 \pi^2}{i}
\dfrac{\mu^{2\varepsilon}
  e^{\gamma \varepsilon}}{(4\pi)^{\varepsilon}} \, \int \dfrac{d^d
  \ell}{(2\pi)^d} \, \dfrac{\ell^{\mu}}{[\ell^2-m_{\tilde g}^2] \, [(\ell+p_{\tilde u_1})^2-m_{u_j}^2] \,
[(\ell+p_{u_2})^2-m_{\tilde u_s}^2]} =p_{\tilde u_1}^{\mu}C_{p_{\tilde u_1}}+p_{u_2}^{\mu}C_{p_{u_2}}
\end{aligned}
\end{equation}
and is given by
\begin{equation}
\begin{aligned}
C_{p_{\tilde u_1}}(m_{\tilde u_1}^2,m_{\tilde \chi_1^0}^2,0;m_{\tilde g}^2,m_{u_j}^2,m_{\tilde u_s}^2)=&\frac{1}{m_{\tilde \chi^0_1}^2-m_{\tilde u_1}^2}\left[B_0(m^2_{\tilde \chi^0_1};m_{u_j}^2,m_{\tilde u_s}^2)-B_0(m^2_{\tilde u_1};m_{\tilde g}^2,m_{u_j}^2) \right.\\
&\left. +\left(m_{\tilde g}^2-m_{\tilde u_s}^2\right)C_0(m^2_{\tilde u_1},m^2_{\tilde \chi^0_1},0;m_{\tilde g}^2,m_{u_j}^2,m_{\tilde u_s}^2)\right]
\end{aligned}
\end{equation}
\medskip

\subsection{Self-energies of quarks and squarks}

In our approximation where we put $m_c=0$, the quark self-energy contribution
with an internal squark and gluino is only needed at $p^2=0$:
\begin{align}
  \Sigma_{f i }^{\tilde g LR}  &=
  \dfrac{\alpha_s}{2\pi}
 W_{fs}^{\tilde q} W_{i + 3,s}^{\tilde q\star} \,
C_F \, m_{\tilde g} \, B_0(0;m_{\tilde{g}}^2,m_{\tilde{q}_s}^2) \,,\\
 \Sigma_{f i }^{\tilde g LL}  &=
   \dfrac{\alpha_s}{2\pi}
 W_{fs}^{\tilde q} W_{i,s}^{\tilde q\star} \,
C_F \, B_1(0;m_{\tilde{g}}^2,m_{\tilde{q}_s}^2)   = - \dfrac{{ \alpha _s
 }}{{4\pi }}   C_F \dfrac{1}{\varepsilon}\delta_{fi}+{\rm finite} \,.
 \label{SigmaLR1}
\end{align}
For the contribution with an internal quark and gluon we get (for arbitrary $p^2$):
\begin{equation}
\renewcommand{\arraystretch}{2}
 \Sigma _{f i }^{\tilde g LL,RR} \left( p^2 \right) =
 \dfrac{\alpha_s}{{4\pi }} {C_F}\left( {d-2} \right){B_1}\left(
       {{p^2};m_{{q_i}}^2,0} \right) \, \delta_{fi} \,,
\end{equation}
\begin{equation}
 \Sigma _{f i }^{\tilde g LR,RL} \left( p^2 \right) =
 \dfrac{\alpha_s}{4\pi} \, C_F \, d \, {m_{{q_i}}} \, {B_0}\left(
       {{p^2};m_{{q_i}}^2,0} \right)\, \delta_{fi} \, .
\end{equation}

\noindent
For the squark self-energies there are three contributions:
\noindent
First, the contribution with internal gluino and quark
\begin{equation}
\begin{aligned}
\Sigma _{{{\tilde u}_s}{{\tilde u}_t}}^{\tilde g}\left( p^2 \right) &= \dfrac{{{\alpha _s}{C_F}}}{\pi
}\left\{ {\left( {W_{js}^{\tilde u*}W_{jt}^{\tilde u} + W_{j + 3,s}^{\tilde u*}W_{j
+ 3,t}^{\tilde u}} \right)\left( {{B_2}\left( {{p^2};m_{\tilde g}^2,m_{{u_j}}^2}
\right) + {p^2}{B_1}\left( {{p^2};m_{\tilde g}^2,m_{{u_j}}^2} \right)} \right)}
\right.\\
&\left.\qquad\qquad {- {m_{\tilde g}}{m_{{u_j}}}\left( {W_{js}^{\tilde
u*}W_{j + 3,t}^{\tilde u} + W_{j + 3,s}^{\tilde u*}W_{jt}^{\tilde u}}
\right){B_0}\left( {{p^2};m_{\tilde g}^2,m_{{u_j}}^2} \right)} \right\} \\ 
                                                                       &=\dfrac{{{\alpha _s}{C_F}}}{\pi
}\dfrac{1}{\varepsilon }\left[ \left( {m_{\tilde g}^2 - \dfrac{{{p^2}}}{2}} \right){\delta _{st}} + \left(
{W_{js}^{\tilde u*}W_{jt}^{\tilde u} + W_{j + 3,s}^{\tilde u*}W_{j + 3,t}^{\tilde
u}}\right)m_{{u_j}}^2 \right.\\ &\left.
\qquad\qquad
- {m_{\tilde g}}{m_{{u_j}}}\left( {W_{js}^{\tilde u*}W_{j + 3,t}^{\tilde u} + W_{j +
3,s}^{\tilde u*}W_{jt}^{\tilde u}} \right) \right] + {\rm finite} \, ,
\label{squark_gluino_SE}
\end{aligned}
\end{equation}

\noindent
second, the contribution with internal squark and gluon
\begin{equation}
\Sigma _{\tilde u_s \tilde u_t }^g \left( {p^2 } \right) = \dfrac{\alpha_s}{{4\pi }}
C_F \left( {2\left( {p^2  + m_{\tilde u_s }^2 } \right)B_0 \left( {p^2 ;m_{\tilde
u_s }^2,0 } \right) - A_0 \left( {m_{\tilde u_s }^2 } \right)} \right)\delta _{st}
\,, 
\label{squark_gluon_SE}
\end{equation}

\noindent
and finally the contribution with a squark tadpole
\begin{equation}
\renewcommand{\arraystretch}{2}
\begin{array}{l}
 \Sigma _{\tilde u_s \tilde u_t }^{\tilde u\tilde u}  = -\dfrac{{\alpha_s}}{{4\pi
}}C_F ( {\delta _{st} A_0 \left( {m_{\tilde u_s }^2 } \right) }\\
\phantom{ \Sigma _{\tilde q_s \tilde q_t }^{\tilde q\tilde q}  =} {-
2\sum\limits_{i,j = 1}^3 {\sum\limits_{s' = 1}^6 {\left( {W_{i + 3s}^{\tilde u\star}
W_{i + 3s'}^{\tilde u} W_{js'}^{\tilde u\star} W_{jt}^{\tilde u}  + W_{is}^{\tilde
u\star} W_{is'}^{\tilde u} W_{j + 3s'}^{\tilde u\star} W_{j + 3t}^{\tilde u} }
\right)A_0 \left( {m_{\tilde u_{s'} }^2 } \right)} } } ) \\ 
 \begin{array}{*{20}c}
   { \phantom{ \Sigma _{\tilde u_s \tilde u_t }^{\tilde u\tilde u}}= -
\dfrac{{\alpha_s}}{{4\pi }} C_F \dfrac{1}{\varepsilon }\left[ {\delta _{st} m_{\tilde u_s }^2 } \right.} \hfill  \\
\phantom{ \Sigma _{\tilde q_s \tilde q_t }^{\tilde q\tilde q}  =} -
2\sum\limits_{i,j = 1}^3 \sum\limits_{s' = 1}^6 \left( {W_{i + 3s}^{\tilde u\star}
W_{i + 3s'}^{\tilde u} W_{js'}^{\tilde u\star} W_{jt}^{\tilde u}  + W_{is}^{\tilde
u\star} W_{is'}^{\tilde u} W_{j + 3s'}^{\tilde u\star} W_{j + 3t}^{\tilde u} }
\right)m_{\tilde u_{s'} }^2  ] + {\rm finite} \, . \\
\end{array}
\end{array} 
\label{squark_squark_SE}
\end{equation}
For further useful information on self-energies and LSZ factors, see
Ref. \cite{Greub:2011ji}.

\medskip

\newpage
\bibliographystyle{hieeetr}
\bibliography{stop-charm-neutralino} 

\end{document}